\documentclass[final,3p,times,twocolumn]{elsarticle}

\usepackage{graphicx}

\usepackage{amsmath,amssymb}

\biboptions{comma,sort&compress}

\journal{Nuc. Phys. (Proc. Suppl.)}

\begin{document}

\begin{frontmatter}



\title{$J^{PC}=1^{++}$ heavy hybrid masses from QCD sum-rules}


\author[label1]{R.T. Kleiv\corref{cor1}}
\author[label2]{D. Harnett}
\author[label1]{T.G. Steele}
\author[label3]{Hong-ying Jin}

\address[label1]{Department of Physics and Engineering Physics, University of Saskatchewan, Saskatoon, SK, S7N 5E2, Canada}
\address[label2]{Department of Physics, University of the Fraser Valley, Abbotsford, BC, V2S 7M8, Canada}
\address[label3]{Zhejiang Institute of Modern Physics, Zhejiang University, Zhejiang Province, P. R. China}

\cortext[cor1]{Speaker}

\begin{abstract}
QCD Laplace sum-rules are used to calculate axial vector $(J^{PC}=1^{++})$ charmonium and bottomonium hybrid masses. Previous sum-rule studies of axial vector heavy quark hybrids did not include the dimension-six gluon condensate, which has been shown to be important in the $1^{--}$ and $0^{-+}$ channels. An updated analysis of axial vector heavy quark hybrids is performed, including the effects of the dimension-six gluon condensate, yielding mass predictions of 5.13 GeV for hybrid charmonium and 11.32 GeV for hybrid bottomonium. The charmonium hybrid mass prediction disfavours a hybrid interpretation of the X(3872), if it has $J^{PC}=1^{++}$, in agreement with the findings of other theoretical approaches. It is noted that QCD sum-rule results for the $1^{--}$, $0^{-+}$ and $1^{++}$ channels are in qualitative agreement with the charmonium hybrid multiplet structure observed in recent lattice calculations. 

\end{abstract}


\end{frontmatter}



\section{Introduction}
\label{theIntroduction} 


Hybrids are mesons that include explicit gluonic degrees of freedom. These can have non-exotic $J^{PC}$ and hence may coexist with heavy quarkonia. The numerous charmonium-like and bottomonium-like ``XYZ'' states discovered since 2003~\cite{Eidelman:2012vu} have inspired the search for hybrids within the charmonium and bottomonium sectors~\cite{Olsen,Olsen2,Godfrey:2008nc,Pakhlova,Close2007}. 

In Ref.~\cite{Harnett:2012gs} QCD Laplace sum-rules were used to perform mass predictions for axial vector $(J^{PC}=1^{++})$ charmonium and bottomonium hybrids. The flux tube model predicts the lightest charmonium hybrids at 4.1-4.2~GeV~\cite{Barnes1995}. Lattice QCD~\cite{Perantonis,Liu:2011rn,Liu:2012ze} yields quenched predictions of about 4.0~GeV for the lightest charmonium hybrids, and unquenched predictions of approximately 4.4~GeV for $1^{++}$ charmonium hybrids in particular. Refs.~\cite{Govaerts:1984hc,Govaerts:1985fx,Govaerts:1986pp} comprise the first studies of heavy quark hybrids using QCD sum-rules. Multiple $J^{PC}$ including $1^{++}$ were examined; however, many of the resulting sum-rules exhibited instabilities, leading to unreliable mass predictions. Refs.~\cite{Qiao:2010zh,Berg:2012gd} re-examined the $1^{--}$ and $0^{-+}$ channels respectively, finding that the dimension-six gluon condensate which was not included in Refs.~\cite{Govaerts:1984hc,Govaerts:1985fx,Govaerts:1986pp} stabilizes the sum-rules in these channels. Motivated by these results, we have investigated the effects of the dimension-six gluon condensate for axial vector heavy quark hybrids using QCD Laplace sum-rules~\cite{Harnett:2012gs}. The resulting mass predictions are discussed with regard to the nature of the X(3872) and in relation to the charmonium hybrid multiplet structure suggested by recent lattice calculations~\cite{Liu:2012ze}.

\section{Laplace Sum-Rules for Axial Vector Heavy Quark Hybrids}
\label{theSumRules} 

The correlation function used to study axial vector ($J^{PC}=1^{++}$) heavy quark hybrids is given by
\begin{gather}
\Pi_{\mu\nu}(q)=i\int d^4x \,e^{i q\cdot x}\langle 0\vert T\left[j_\mu(x)j_\nu(0)\right]\vert 0\rangle
\label{basic_corr}
\\
j_\mu=\frac{g}{2}\bar Q\lambda^a\gamma^\nu\tilde G^a_{\mu\nu}Q\,,~\tilde G^a_{\mu\nu}=\frac{1}{2}\epsilon_{\mu\nu\alpha\beta}G^a_{\alpha\beta}\,,
\label{current}
\end{gather} 
with $Q$ representing a heavy quark field~\cite{Govaerts:1985fx}. The transverse part $\Pi_{\rm V}$ of \eqref{basic_corr} couples to $1^{++}$ states
\begin{equation}
\Pi_{\mu\nu}(q)=\left(\frac{q_\mu q_\nu}{q^2}-g_{\mu\nu} \right)\Pi_{\rm V}(q^2)+\frac{q_\mu q_\nu}{q^2}\Pi_{\rm S}(q^2)~.
\label{corr_tensor}
\end{equation}

\begin{figure}[hbt]
\centering
\includegraphics[scale=0.45]{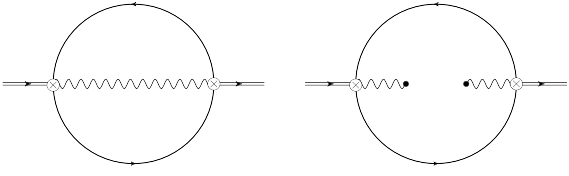}
\caption{Feynman diagram for the leading-order perturbative and $\langle \alpha \, G^2\rangle$ contributions to $\Pi_{\rm V}$. The current is represented by the $\otimes$ symbol.}
\label{pert_g2_fig}
\end{figure}

In Refs.~\cite{Govaerts:1985fx,Govaerts:1984hc} the perturbative and gluon condensate $\langle \alpha \, G^2\rangle=\langle \alpha G^a_{\mu\nu} G^a_{\mu\nu}\rangle$ contributions to the imaginary part of $\Pi_{\rm V}(q^2)$ were calculated to leading order. The Feynman diagrams for these are represented in Fig.~\ref{pert_g2_fig}. 

For brevity only imaginary parts of the perturbative and $\langle \alpha \, G^2\rangle$ contributions are given; the full expressions may be found in Ref~\cite{Harnett:2012gs}. We find 
\begin{gather}
\begin{split}
{\rm Im}\Pi_{\rm V}^{\rm pert}(q^2)=&
\frac{\alpha m^6}{180\pi^2z^2}
\Biggl[
\left(15-35z-22z^2-216z^3\right.
\\ &\left.+48z^4\right)\sqrt{z-1}\sqrt{z} 
+15\left(1-3z+16z^3\right)
\\
&\times \Biggl. \log\left[\sqrt{z-1}+\!\sqrt{z}\right]
\Biggr] \,,
\label{Im_Pi_pert}
\end{split}
\\
\begin{split}
{\rm Im}\Pi^{\rm GG}_{\rm V}(q^2)&=-\frac{ m^2\langle \alpha G^2\rangle}{18}\left(1+2z\right)\frac{\sqrt{z-1}}{\sqrt{z}}\,,
\\
&z=\frac{q^2}{4m^2}\,, \quad z>1 \,.
\label{Im_Pi_GG} 
\end{split}
\end{gather}
Expressions \eqref{Im_Pi_pert} and \eqref{Im_Pi_GG} are in complete agreement with the corresponding integral representations given in~\cite{Govaerts:1985fx,Govaerts:1984hc}.

\begin{figure}[hbt]
\centering
\includegraphics[scale=0.45]{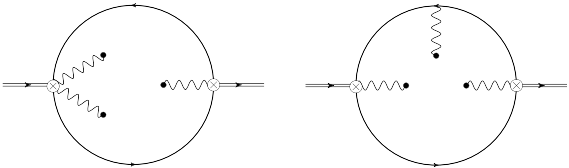}
\caption{Feynman diagram for the leading-order $\langle g^3 G^3\rangle$ contribution to $\Pi_{\rm V}$. Additional diagrams related by symmetry are not shown.
}
\label{GGG_fig}
\end{figure}

The dimension-six gluon condensate $\langle g^3 G^3\rangle=\langle g^3 f_{abc} G^a_{\mu\nu} G^b_{\nu\alpha} G^c_{\alpha\mu}\rangle$ contributions are now determined. These were not calculated in Refs.~\cite{Govaerts:1985fx,Govaerts:1984hc}, and are represented by the diagrams in Fig.~\ref{GGG_fig}. The full expression for these contributions is

\begin{gather}
\begin{split}
\Pi_{\rm V}^{\rm GGG}(q^2)=&\frac{\langle g^3G^3\rangle}{1152\pi^2}
\Biggl[
\frac{3(17z-9)}{z-1}-\frac{3(17-46z+27z^2)}{(z-1)^2}
\Biggr. 
\\&+\left(\frac{2z(2-9z+6z^2)}{(z-1)^2}-\frac{4z\left(3z-1\right)}{z-1}\right)
\\
&\Biggl.\times\,\phantom{}_2F_1\left(1, 1; 5/2;z\right)
\Biggr]\,.
\end{split}
\label{Pi_GGG}
\end{gather}
The imaginary part of \eqref{Pi_GGG} is
\begin{gather}
\begin{split}
{\rm Im}\Pi_{\rm V}^{\rm GGG}(q^2)=&\frac{\langle g^3G^3\rangle}{384\pi }\frac{\sqrt{z-1}}{\sqrt{z}}\left[
\frac{2(1-3z)}{z-1} \right.
\\
&\left. +\frac{(2-9z+6z^2)}{(z-1)^2} \right] \,,\quad z>1\,,
\label{Im_Pi_GGG}
\end{split}
\end{gather} 
which is singular at $z=1$. This poses a problem since the sum-rules will involve integrating \eqref{Im_Pi_GGG} from $z=1$. Below it will be shown how this difficulty can be overcome.

We now formulate the QCD Laplace sum-rules~\cite{Shifman:1978bx,Shifman:1978by}. Using a resonance plus continuum model for the hadronic spectral function
\begin{gather}
\rho(t) = \rho^{\rm had}(t) + \theta(t-s_0) \rm{Im}\Pi^{QCD}(t)\,,
\end{gather}
the Laplace sum-rules are given by
\begin{gather}
{\cal L}_{k}^{\rm QCD}\left(\tau,s_0\right)  = \frac{1}{\pi}\int_{t_0}^{\infty} t^k
   \exp\left[ -t\tau\right] \rho^{\rm had}(t)\; dt \,,
\label{final_laplace}
\end{gather}
where $t_0$ is the hadronic threshold. The quantity on the left hand side of \eqref{final_laplace} is given by
\begin{gather}
\begin{split}
{\cal L}_k^{\rm QCD}\left(\tau,s_0\right)=&\frac{1}{\tau}\hat B\left[\left(-1\right)^k Q^{2k}\Pi_{\rm V}\left(Q^2\right)\right] 
\\ 
&-\frac{1}{\pi}  \int_{s_0}^{\infty} t^k \exp \left[-t\tau  \right]  {\rm Im} \Pi_{\rm V}(t)\; dt \,,
\label{laplace} 
\end{split}
\end{gather}
where $Q^2=-q^2$ and $s_0$ is the continuum threshold. $\hat B$~is the Borel transform, which is closely related to the inverse Laplace transform~\cite{Bertlmann:1984ih}. The singular terms in \eqref{Pi_GGG} that are irrelevant for \eqref{Im_Pi_GGG} are, however, relevant for the inverse Laplace transform. It is the inclusion of these terms that allows the integration of \eqref{Im_Pi_GGG} from $z=1$ to be defined as a limiting procedure. Thus the imaginary part \eqref{Im_Pi_GGG} is insufficient to formulate the sum-rules for $1^{++}$ hybrids, as found for $0^{-+}$ hybrids~\cite{Berg:2012gd}.

From the results for the leading order perturbative~\eqref{Im_Pi_pert}, $\langle \alpha G^2 \rangle$~\eqref{Im_Pi_GG}, $\langle g^3 G^3 \rangle$~\eqref{Pi_GGG}~and~\eqref{Im_Pi_GGG} contributions, we find
\begin{gather}
\begin{split}
{\cal L}_0^{\rm QCD}\left(\tau,s_0\right)=&\frac{4m^2}{\pi}
\Biggl[
\int_1^{s_0/4m^2} 
\left[
{\rm Im}\Pi_{\rm V}^{\rm pert}\left(4m^2 x\right) 
\right. 
\Biggr.
\\
&\left.
+{\rm Im}\Pi_{\rm V}^{\rm GG}\left(4m^2 x\right)
\right]
\exp{\left(-4m^2\tau x\right)\,dx}
\\
&+\lim_{\eta\to 0^+}
\left(
\int_{1+\eta}^{s_0/4m^2} {\rm Im}\Pi_{\rm V}^{\rm GGG}(4m^2 x)
\right.
\\
&\times \exp{\left(-4m^2\tau x\right)\,dx}
\\
&+\left.
\Biggl.
\frac{4m^2\langle g^3G^3\rangle}{192\pi^2\sqrt{\eta}}\exp{(-4m^2\tau)}
\right)
\Biggr]
\,,
\end{split}
\label{L_0}
\\
{\cal L}_1^{\rm QCD}\left(\tau,s_0\right)=-\frac{\partial}{\partial\tau}{\cal L}_0^{\rm QCD}\left(\tau,s_0\right)\,.
\label{L_1}
\end{gather}
The mass and coupling are functions of the renormalization scale $\mu$ in the $\overline{\rm MS}$-scheme. After evaluating the $\tau$ derivative in \eqref{L_1}, renormalization group improvement may be implemented by setting $\mu=1/\sqrt{\tau}$~\cite{Narison:1981ts}.

\section{Analysis: Mass Predictions for Axial Vector Heavy Quark Hybrids}  
\label{theAnalysis}

To make mass predictions for axial vector heavy quark hybrids we utilize a single narrow resonance model
\begin{equation}
 \frac{1}{\pi}\rho^{\rm had}(t)=f^2\delta\left(t-M^2\right)\,.
 \label{narrow_res}
\end{equation}
Inserting \eqref{narrow_res} in \eqref{final_laplace} gives
\begin{equation}
{\cal L}_k^{\rm QCD}\left(\tau,s_0\right)=f^2 M^{2k}\exp{\left(-M^2\tau\right)}\,,
\label{narrow_sr}
\end{equation}
which can be used to calculate the ground state mass $M$ via the ratio
\begin{equation}
M^2=\frac{{\cal L}_1^{\rm QCD}\left(\tau,s_0\right)}{{\cal L}_0^{\rm QCD}\left(\tau,s_0\right)}\,.
\label{ratio}
\end{equation}
Before using \eqref{ratio} to calculate the mass, the QCD input parameters must be specified. For the charmonium and bottomonium hybrid analyses we use one-loop $\overline{{\rm MS}}$ expressions for the coupling and quark masses:
\begin{gather}
\begin{split}
&\alpha(\mu)=\frac{\alpha\left(M_\tau\right)}{1+\frac{25\alpha\left(M_\tau\right)}{12\pi}\log{\left(\frac{\mu^2}{M_\tau^2}\right)}} \,, \;
m_c(\mu)=\overline m_c\left(\frac{\alpha(\mu)}{\alpha\left(\overline m_c\right)}\right)^\frac{12}{25} \,;
\\
&\alpha(\mu)=\frac{\alpha\left(M_Z\right)}{1+\frac{23\alpha\left(M_Z\right)}{12\pi}\log{\left(\frac{\mu^2}{M_Z^2}\right)}} \,, \;
m_b(\mu)=\overline m_b\left(\frac{\alpha(\mu)}{\alpha\left(\overline m_b\right) }\right)^\frac{12}{23} \,.
\end{split}
\end{gather}
The numerical values of the QCD parameters are given in Table~\ref{QCD_parameters}.
\begin{table}[ht]
\centering
  \begin{tabular}{|| l | l ||}
  \hline
  \hline
  $\alpha\left(M_\tau\right)$ 		& $0.33$ 								\\
  $\overline{m}_c$			& $\left(1.28 \pm 0.02\right) {\rm GeV}$ 				\\
  $\alpha\left(M_Z\right)$		& $0.118$								\\
  $\overline{m}_b$			& $\left(4.17 \pm 0.02\right) {\rm GeV}$ 				\\
  $\langle \alpha \, G^2 \rangle$	& $\left(7.5 \pm 2.0\right)\times 10^{-2}{\rm GeV}^4$			\\
  $\langle g^3G^3\rangle$ 		& $\left(8.2\pm 1.0\right){\rm GeV^2}\langle \alpha \, G^2\rangle$	\\
  \hline
  \hline
  \end{tabular}
\caption{QCD parameters. The quark masses, $M_\tau$ and $M_Z$ are taken from Ref.~\cite{pdg}. The values of $\alpha\left(M_\tau\right)$ and $\alpha\left(M_Z\right)$ are from Ref.~\cite{Bethke:2009jm}. Numerical values of the condensates are taken from Ref.~\cite{Narison:2010cg}.}
\label{QCD_parameters}
\end{table}

The sum-rule window is determined following Ref.~\cite{Shifman:1978bx} by requiring that contributions from the continuum are less than 30\% of total and non-perturbative contributions are less than 15\% of total. These criteria are then used to constrain the Borel parameter $\tau$. This leads to the sum-rule windows of $5.3\,{\rm GeV^2} < 1/\tau <7.3\,{\rm GeV^2}$ for the charmonium hybrid and $7.8\,{\rm GeV^2} < 1/\tau <25.0\,{\rm GeV^2}$ for the bottomonium hybrid. 

\begin{figure}[hbt]
\centering
\includegraphics[scale=0.85]{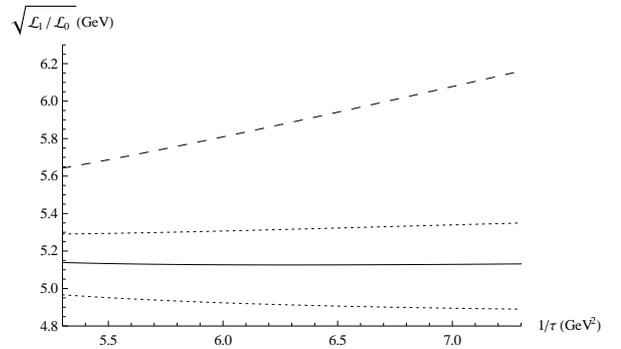}
\caption{The ratio ${\cal L}_1^{\rm QCD}\left(\tau,s_0\right)/{\cal L}_0^{\rm QCD}\left(\tau,s_0\right)$
for hybrid charmonium is shown as a function of the Borel scale $1/\tau$ for the optimized value $s_0=33\,{\rm GeV^2}$ (solid curve). The ratio is also shown for $s_0=38\,{\rm GeV^2}$ (upper dotted curve), $s_0=28\,{\rm GeV^2}$ (lower dotted curve) and $s_0\to\infty$ (uppermost dashed curve). Central values of the QCD parameters have been used.}
\label{charm_opt}
\end{figure}

\begin{figure}[hbt]
\centering
\includegraphics[scale=0.85]{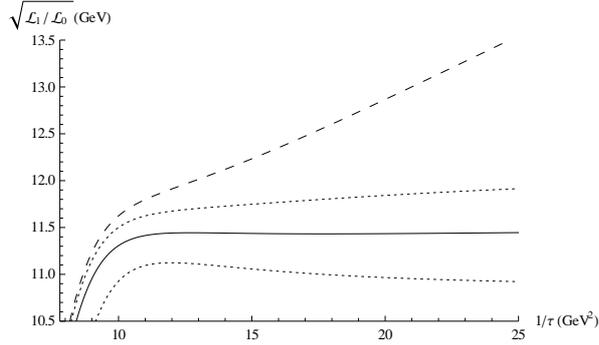}
\caption{The ratio ${\cal L}_1^{\rm QCD}\left(\tau,s_0\right)/{\cal L}_0^{\rm QCD}\left(\tau,s_0\right)$ for hybrid bottomonium is shown  as a function of the Borel scale $1/\tau$ for the optimized value $s_0=150\,{\rm GeV^2}$ (solid curve). For comparison the ratio is also shown for $s_0=170\,{\rm GeV^2}$ (upper dotted curve), $s_0=130\,{\rm GeV^2}$ (lower dotted curve) and  $s_0\to\infty$ (uppermost dashed curve). Central values of the QCD parameters have been used.}
\label{bottom_opt}
\end{figure}

The continuum threshold $s_0$ is optimized by first determining the smallest value of $s_0$ for which the ratio \eqref{ratio} stabilizes (exhibits a minimum) within the respective sum-rule windows. Then the optimal value is fixed by the $s_0$ which has the best fit to a constant within the sum-rule window. The mass prediction \eqref{ratio} is shown for hybrid charmonium in Fig.~\ref{charm_opt} and for hybrid bottomonium in Fig.~\ref{bottom_opt}.

We predict the masses of axial vector charmonium and bottomonium hybrids to be $5.13\pm0.25\,{\rm GeV}$ and $11.32\pm0.32\,{\rm GeV}$, respectively. The uncertainties are due to the QCD parameter uncertainties given in Table~\ref{QCD_parameters} and are dominated by variations in $\langle \alpha \, G^2 \rangle$. This is in contrast to the pseudoscalar where variations in $\langle g^3 G^3 \rangle$ dominate~\cite{Berg:2012gd}. These predictions for axial vector heavy quark hybrids are in agreement with those of Refs.~\cite{Govaerts:1984hc,Govaerts:1986pp}, suggesting that the effects of $\langle g^3 G^3 \rangle$ are less important for the $1^{++}$ channel than for the $1^{--}$ and $0^{-+}$ channels.

\section{Conclusions}
\label{theConclusion}

In Ref.~\cite{Harnett:2012gs} we have studied $J^{PC}=1^{++}$ heavy quark hybrids using QCD sum-rules. For the first time we have calculated the contributions from $\langle g^3 G^3 \rangle$, which were found to be important for the $1^{--}$~\cite{Qiao:2010zh} and $0^{-+}$~\cite{Berg:2012gd} channels. We find that $\langle g^3 G^3 \rangle$ has less effect on the $1^{++}$ channel, resulting in mass predictions of $5.13\pm0.25\,{\rm GeV}$ for hybrid charmonium and $11.32\pm0.32\,{\rm GeV}$ for hybrid bottomonium, in agreement with the range of predictions given in Refs.~\cite{Govaerts:1984hc,Govaerts:1986pp}.

The X(3872) has possible $J^{PC}$ assignments of $1^{++}$ or $2^{-+}$~\cite{Abulencia:2006ma,Abe:2005iya}, but $1^{++}$ is strongly favoured~\cite{Brambilla:2010cs}. In Ref.~\cite{Li:2004sta} it was suggested that the X(3872) is a hybrid, but this interpretation has been largely ruled out since the flux-tube~model~\cite{Barnes1995} and lattice QCD~\cite{Perantonis,Liu:2011rn,Liu:2012ze} predict that the lightest charmonium hybrids have masses significantly greater than that of the X(3872). If it is shown to have $J^{PC}=1^{++}$, our mass prediction of $5.13\,\rm{ GeV}$ is in agreement with the results of other theoretical approaches that disfavour a charmonium hybrid interpretation of the X(3872).

In Ref.~\cite{Liu:2012ze} it is suggested that $0^{-+}$ and $1^{--}$ are members of a ground state charmonium hybrid multiplet, while $1^{++}$ is a member of a multiplet of excited charmonium hybrids. The present result and those of Refs.~\cite{Qiao:2010zh,Berg:2012gd} seem to be in qualitative agreement with this multiplet structure, although the mass splittings are significantly larger than those of Ref.~\cite{Liu:2012ze}. Future work to update remaining unstable sum-rule channels in Refs.~\cite{Govaerts:1985fx,Govaerts:1984hc,Govaerts:1986pp} to include the effects of $\langle g^3 G^3 \rangle$ would clarify the QCD sum-rule predictions for the spectrum of charmonium hybrids.

\bigskip
\noindent
{\bf Acknowledgements:}  We are grateful for financial support from the Natural Sciences and Engineering Research Council of Canada (NSERC).













\end{document}